\documentclass[12pt,a4paper]{article}
\usepackage{amsmath,amsfonts,amsthm,amssymb,mathrsfs,mathtools}
\usepackage{times}
\usepackage{empheq}
\usepackage{authblk}
\usepackage{enumitem}
\usepackage{hyperref}
\usepackage[mathcal]{euscript}
\usepackage{xcolor}

\numberwithin{equation}{section}

\theoremstyle{plain}
\newtheorem{thm}{Theorem}[section]
\newtheorem*{smp*}{Strong Maximum Principle}
\newtheorem{lem}[thm]{Lemma}
\newtheorem{cor}[thm]{Corollary}
\newtheorem{prop}[thm]{Proposition}
\newtheorem{claim}{Claim}

\newtheorem*{claim*}{Claim}

\theoremstyle{definition}

\theoremstyle{remark}
\newtheorem{remark}{Remark}[section]
\newtheorem*{remarks}{Remarks}

\newcommand{\da}{\delta}
\newcommand{\Da}{\Delta}

\newcommand{\la}{\lambda}
\newcommand{\Oa}{\Omega}

\newcommand{\vn}{\varepsilon}

\newcommand{\pl}{\partial}
\newcommand{\fc}{\frac}
\newcommand{\na}{\nabla}

\newcommand{\lt}{\left}
\newcommand{\rt}{\right}

\newcommand{\wto}{\rightharpoonup}

\newcommand{\mf}{\mathbf}
\newcommand{\mb}{\mathbb}
\newcommand{\ml}{\mathcal}

\newcommand{\up}{\textup}

\newcommand{\wt}{\widetilde}
\newcommand{\dx}{\,d\mf x}
\begin{document}

\fontsize{12}{16pt plus.4pt minus.3pt}\selectfont

\title{Phase transition between two-component and three-component ground states of spin-1 Bose-Einstein condensates
}
\author[,1]{Liren Lin\footnote{lirenlin2017@gmail.com}}
\author[,2]{I-Liang Chern\footnote{chern@math.ntu.edu.tw}}
\affil[1]{Department of Applied Mathematics, National Sun Yat-sen University, Taiwan}
\affil[2]{Department of Mathematics, National Taiwan University, Taiwan}
\date{}

\maketitle

\begin{abstract}
For an antiferromagnetic spin-1 Bose-Einstein condensate under 
an applied uniform magnetic field, its ground state 
$(\psi_1,\psi_0,\psi_{-1})$ undergoes a phase transition from a two-component state 
($\psi_0 \equiv 0$) to a three-component state ($\psi_j\ne 0$ for all $j$)
at a critical value of the magnetic field. 
This phenomenon has been observed in numerical 
simulations as well as in experiments. In this paper, we provide a 
mathematical proof based on 
a simple principle found by the authors:
a redistribution of the mass densities between different 
components will decrease the kinetic energy.
\end{abstract}

\tableofcontents

%
%
%
%
%
%
%
%

\section{Introduction}

A Bose-Einstein condensate (BEC) is a state of matter that is 
formed when a dilute gas of bosons is cooled to near absolute zero. 
When a BEC is confined by an optical trap, all its hyperfine spin 
states can be active. In the mean-field theory, a spin-$f$ BEC is described by a 
$(2f+1)$-component complex functions $\Psi = (\psi_{f},\psi_{f-1},...,\psi_{-f})$.
Since the first realization of such 
spinor BECs \cite{Stamper1998.3}, their rich structures have 
attracted a lot of attention \cite{Ohmi1998,Ho1998,Law1998.12,Stenger1998.11,Zhang2003,
Kawaguchi_Ueda,Bao2008.5,Cao2011}. 

For an antiferromagnetic 
spin-1 BEC $(\psi_1,\psi_0,\psi_{-1})$ under a weak applied uniform magnetic field, its ground state is a 
two-component state (2C state) with $\psi_0 \equiv 0$. 
As the strength of the applied magnetic field increases, the 
ground state undergoes a phase 
transition to a three-component state (3C state), i.e. $\psi_j\ne 0$ for all $j=1,0,-1$,
at a critical value of the applied magnetic field. 
This phenomenon has 
been known from numerical simulations for a long time
\cite{Lim2008.12, chen2015}, and was also 
observed experimentally \cite{expphase}. 
In this paper, we provide a mathematical proof of its occurrence based 
on a simple principle found by the authors in 
\cite{LinChern2011}, which says that a redistribution between the mass densities 
$|\psi_j|^2$ will not increase the kinetic energy. 
Using this technique in \cite{LinChern2011}, we have shown
that the ground state for an antiferromagnetic system is a 2C state 
when there is no applied magnetic field.  
In this paper, we push further by showing some fundamental properties
in the presence of a magnetic field, particularly in the 2C-3C phase 
transition phenomenon.

\subsection{Mean-field model}

In the mean-field approximation, a spin-1 BEC system 
$\Psi(x) = (\psi_1(x),\psi_0(x),\psi_{-1}(x))$ ($x\in\mb R^3$)
in a uniform magnetic field $B$ is described by the energy functional\footnote{To 
save notation, we write $\int $ instead of $\int_{\mb{R}^3}$ to denote 
an integration over the entire $\mb R^3$ space.
} \cite{Zhang2003, Stenger1998.11,Ohmi1998}
\begin{align*}
\ml E[\Psi] 
&= \int\Big\{H_{kin}(\Psi)+H_{pot}(\Psi)+H_{n}(\Psi)+H_{s}(\Psi)+H_{Zee}(\Psi)\Big\}\dx
\\
&\coloneqq\int\Big\{
\frac{\hbar^2}{2m_a} \sum_{j=-1}^1 |\nabla\psi_j|^2 + V(x)|\Psi|^2 + \frac{c_n}{2} |\Psi|^4 
+ \frac{c_s}{2} |\Psi^\dag {\mf F} \Psi|^2 +\sum_{j=-1}^1 E_j|\psi_j|^2\Big\}\,d x.
\end{align*}
$H_{kin}(\Psi)$ is the kinetic energy term, 
where $\hbar$ is the reduced Planck constant, and 
$m_a$ is the atomic mass.
The function $V( x)$ in $H_{pot}(\Psi)$ represents a trap potential.
$H_n(\Psi)$ describes the spin-independent interaction between atoms.
We will consider $c_n>0$, meaning that the interaction is repulsive.
$H_s(\Psi)$ describes the spin-dependent interaction, 
in which $\mf F= (F_x,F_y,F_z)$ is a vector of Hermitian matrices given by
\[
F_x = \frac{1}{\sqrt{2}}\lt(\begin{array}{ccc}0&1&0\\1&0&1\\0&1&0\end{array}\rt),
F_y = \frac{i}{\sqrt{2}}\lt(\begin{array}{ccc}0&-1&0\\1&0&-1\\0&1&0\end{array}\rt),
F_z = \lt(\begin{array}{ccc}1&0&0\\0&0&0\\0&0&-1\end{array}\rt),
\]
and $\Psi^\dag\mf F\Psi $ denotes the vector 
$(\Psi^\dag  F_x\Psi,\Psi^\dag  F_y\Psi,\Psi^\dag  F_z\Psi)\in \mathbb R^3$.
We will consider $c_s>0$ (anti-ferromagnetic interaction), for which 
a typical example is $^{23}$Na. 
Finally, $H_{Zee}(\Psi)$
comes from the interaction of spinor atoms with the applied magnetic field 
$B$. The number $E_j$ (which depends on $B$) is called the Zeeman shift 
associated with the spin component $j$.

The system has two conserved quantities:
\begin{itemize}
\item Total number of atoms: $\ml N[\Psi] \coloneqq \int |\Psi|^2\, d x= N$,
\item Magnetization: $\ml M[\Psi] \coloneqq \int (|\psi_1|^2 - |\psi_{-1}|^2)\, d x=M$,
\end{itemize}
where $N$ and $M$ are two constants, $N>0$ and $|M|\le N$.
A ground state is a minimizer $\Psi$ of the energy functional $ \ml E$ subject 
to these two constraints. 

Note that we can write
\begin{align}\label{ezpq}
\int H_{Zee}(\Psi)\,d x &= \int \Big\{E_0|\Psi|^2 - p(|\psi_1|^2-|\psi_{-1}|^2) 
+ q(|\psi_1|^2+|\psi_{-1}|^2)\Big\}\,d x\\
& = E_0 N - p M + q\int (|\psi_1|^2+|\psi_{-1}|^2)\,d x,
\end{align}
where $p=\frac{1}{2}(E_{-1}-E_1)$ and 
$q = \frac{1}{2}(E_1+E_{-1}-2E_0)$. Physically, we have 
$p\sim B$ and $q\sim B^2$, and hence $p$ and $q$
are called the linear and 
quadratic Zeeman energies, respectively.
Due to the conservation of $N$ and $M$,
the values of $E_0$ and $p$ has no influence on the ground-state formations,
and the only relevant parameter for our study is $q$.
Our main goal is hence to prove that there exists a critical 
value of $q$ at which the phase transition occurs. 

\subsection{Reduced model for ground states}\label{redmodel}

Let us express $\psi_j( x) = u_j( x) e^{i\theta_j( x)}$ 
with $u_j \ge 0$ 
and $\theta_j \in\mb R$.
For our purpose, it is possible to 
consider a reduced model on the amplitude vector 
\begin{align*}
\mf{u}=(u_1,u_0,u_{-1})=(|\psi_1|,|\psi_0|,|\psi_{-1}|).
\end{align*}
To see it, note that the phase functions $\theta_j$
only appear in $H_{kin}(\Psi)$ and $H_s(\Psi)$, which by 
direct computation are written as 
\begin{align*}
H_{kin}(\Psi) &= 
\frac{\hbar^2}{2m_a} \sum_{j=-1}^1\Big\{
 |\nabla u_j|^2 + u_j^2|\nabla \theta_j|^2\Big\}\\
H_s(\Psi) &=
\frac{c_s}{2} \Big\{2u_0^2 \big(u_1^2+u_{-1}^2 + 2 u_1 u_{-1}
\cos\Delta\theta\big)+(u_1^2-u_{-1}^2)^2 \Big\},
\end{align*}
where $\Delta\theta = \theta_1+\theta_{-1}-2\theta_0$.
From these expressions (and recall that we consider $c_s>0$), when 
least energy is achieved, all the $\theta_j$'s must be constants 
(so that $|\nabla\theta_j|=0$) and satisfy $\cos \Delta\theta = -1$
(so that $u_1^2+u_{-1}^2 + 2 u_1 u_{-1}
\cos\Delta \theta = (u_1-u_{-1})^2$).

Besides the above observation, recall that the values 
of $E_0$ and $p$ in the expression \eqref{ezpq} for 
$H_{Zee}$ will not influence the ground states. 
Hence we will ignore the corresponding terms by setting $E_0=p=0$, and thus
\begin{align*}
H_{Zee}(\Psi) = q(|\psi_1|^2+|\psi_{-1}|^2) = q(u_1^2+u_{-1}^2). 
\end{align*}
Moreover, without loss of generality, we shall simplify the notation by i) ignoring 
the coefficient $\hbar^2/2m_a$ of $H_{kin}$, ii) setting the number of atoms $N$ to be 
$1$, and iii) omitting the factor $1/2$ from $c_n$ and $c_s$. To be precise, 
i) and ii) can be 
achieved by a suitable normalization, i.e. letting $\mf u( x) = |a\Psi(b x)|$ for 
some constants $a,b>0$ instead of $\mf u( x)=|\Psi( x)|$, while iii) may be 
regarded just as a redefinition of $c_n$ and $c_s$.

In summary, the mathematical model we are going to study reduces to the 
following energy functional for $\mf u=(u_1,u_0,u_{-1})$:
\begin{align*}
\ml E[\mf{u}]  = \ml{E}_{kin}[\mf{u}]+\ml{E}_{pot}[\mf{u}]+\ml{E}_n[\mf{u}]
+\ml{E}_s[\mf{u}]+\ml{E}_{Zee}[\mf{u}],
\end{align*}
where
\begin{align*}
\ml{E}_{kin}[\mf{u}] 
&= \int H_{kin}(\mf{u})\,d x \coloneqq \int \sum_j |\na u_j|^2\,d x
\\
\ml{E}_{pot}[\mf{u}] 
&= \int H_{pot}(\mf{u})\,d x \coloneqq \int V( x)|\mf{u}|^2\,d x
\\
\ml{E}_n[\mf{u}] &= \int H_n(\mf{u})\,d x \coloneqq \int c_n |\mf{u}|^4\,d x
\\
\ml{E}_s[\mf{u}] &= \int H_s(\mf{u})\,d x 
\coloneqq \int c_s\big[2u_0^2(u_1-u_{-1})^2+(u_1^2-u_{-1}^2)^2\big]\,d x
\\
\ml{E}_{Zee}[\mf{u}] &= \int H_{Zee}(\mf{u})\,d x \coloneqq \int q(u_1^2+u_{-1}^2)\,d x,
\end{align*}
with constraints:
\begin{align*}
\ml{N}[\mf u] &\coloneqq \int |\mf u|^2\,d x = 1\\
\ml {M}[\mf u] &\coloneqq \int (u_1^2 - u_{-1}^2)\,d x =M.
\end{align*}
We will also call a minimizer $\mf u$ of $\ml E$ subject to the two constraints
a ground state. 

\subsubsection{Assumptions}

We consider the following setting. 
\begin{enumerate}[label={(A\arabic*)}]
\item \label{A1}
$V\in L^{\infty}_{loc}(\mb{R}^3)$, $V\ge 0$ and $V(x)\to +\infty$  
as $|x|\to\infty$.
\item \label{A2}
$c_n > 0$ and $c_s > 0$. 
\item \label{A3}
$q\ge 0$.
\item \label{A4}
$0\le M\le 1$.
\end{enumerate}

\begin{remarks}~
\begin{itemize}
\item In laboratory, a common trap potential $V(x)$ is a quadratic function, 
which satisfies \ref{A1}. The assertion $V(x)\to +\infty$ can be stated more
precisely as 
\begin{align*}
\{x\,|\, V(x)\le \xi\}\mbox{  is a bounded set in }\mb R^3\mbox{ for all }\xi>0.
\end{align*}
This guarantees that $V(x)$ traps the 
repulsive system mostly in localized regions. 
\item For \ref{A4}, we have mentioned that $|M|\le N = 1$. Due to the symmetry of the 
roles of $u_1$ and $u_{-1}$, it suffices to consider nonnegative $M$. 
\end{itemize}
\end{remarks}

\subsection{Main theorem}

We will study ground-state formations under different values 
of $M$ and $q$. 
Recall that by a 2C state we mean a state $\mf u$ with 
$u_0\equiv 0$ and $u_1,u_{-1}\ne 0$, and a 3C state 
is one that with $u_j\ne 0$ for all $j=1,0,-1$.
Our main theorem on the 2C-3C phase transition phenomenon
is the following.

\begin{thm}\label{thm:qc}
For fixed $0<M<1$, there is a positive number $q_c(M)$ such that for 
$0 \le q < q_c(M)$, there is a unique 2C ground state, 
while for $q>q_c(M)$, any ground state is a 3C state.
\end{thm}

We recommend the reader to consult Figure 5 of \cite{Lim2008.12} 
for a clear phase diagram on the $(q, M)$-plane
depicting the phenomenon. 

The organization of the rest of the paper is as follows:
In Section \ref{preli}, we give some fundamental facts 
such as the existence of ground states.
In Section \ref{sfpro} we recap the idea of mass redistribution 
and prove some more properties
about ground states that will be needed. 
The main theorem is proved in Section \ref{sec:ptp}. After that, some additional 
issues are discussed. 

%
%
%
%
%
%

\section{Preliminaries}\label{preli}

\subsection{Notation}

We begin our study of the model introduced in Section
\ref{redmodel}.
To facilitate later discussions, we define some more notation.
Let $\mb{B}$ be the class of all triples 
$\mf{u}=(u_1,u_0,u_{-1}):\mb R^3\to\mb R^3$, where
\begin{align*}
u_j\in H^1(\mb R^3)\cap L^2_V(\mb R^3) \cap L^4(\mb R^3).
\end{align*}
Here $L_V^2(\mb R^3)$ denotes the $V$-weighted $L^2$ space:
\begin{align*}
f\in L_V^2(\mb R^3)\quad \mbox{if} \quad 
\|f\|_{L_V^2}^2 \coloneqq \int V(x)|f|^2\,dx <\infty.
\end{align*}
$\mb{B}$ is a Banach space with norm
\begin{align}\label{normofB}
\|\mf{u}\|_{\mb{B}} \coloneqq 
\sum_{j=-1}^1\big(\|u_j\|_{H^1}+\|u_j\|_{L_V^2}+\|u_j\|_{L^4}\big).
\end{align}
The admissible class on which we are going to minimize 
the energy functional $\ml{E}$ is
\begin{align*}
\mb{A}_M \coloneqq \lt\{
\mf{u}\in\mb{B} \,|\, u_j\ge 0\mbox{ for each }j,\ \ml{N}[\mf{u}]=1\,\up{ and }\,
\ml{M}[\mf{u}]=M
\rt\}.
\end{align*}
The ground-state energy is
\begin{align*}
E_g = E_g(M,q) \coloneqq \inf_{\mf{v}\in\mb{A}_M}\ml{E}[\mf{v}].
\end{align*}
$\mf u\in\mb A_M$ is called a ground state if $\ml E[\mf u]=E_g$.
We will use $\mb{G}_{M,q}$ to denote the set of ground state(s)
for given $M,q$. At times when a discussion concerns different values of $q$,
we will use $\ml{E}[\mf u,q]$ instead of $\ml{E}[\mf u]$ to indicate precisely the 
value of $q$ being considered.

\subsection{Basic properties}

We collect some basic facts about ground states in the following theorem.

\begin{thm}\label{thm:exist}
Under the assumptions \ref{A1} $\sim$ \ref{A4}, the ground state set $\mb{G}_{M,q}\ne\emptyset$. 
If $\mf u\in\mb{G}_{M,q}$, then $\mf{u}$
satisfies the following Euler-Lagrange system:
\begin{empheq}[left=\empheqlbrace]{align*}
(\mu + \la) u_1  &= \ml{L}u_1+2c_s\lt[
u_0^2(u_1-u_{-1})+u_1(u_1^2-u_{-1}^2)\rt]+qu_1
\tag{2.2a}\label{2.2a}
\\
\mu u_0  &= \ml{L}u_0 + 2c_s u_0 (u_1-u_{-1})^2
\tag{2.2b}\label{2.2b}
\\
(\mu - \la) u_{-1}  &= \ml{L}u_{-1}+2c_s\lt[
u_0^2(u_{-1}-u_1)+u_{-1}(u_{-1}^2-u_1^2)\rt]+qu_{-1},
\tag{2.2c}\label{2.2c}
\end{empheq}
where $\ml{L}=-\Da + V + 2c_n |\mf{u}|^2$, and $\la,\mu$ 
are Lagrange multipliers induced by the constraints 
$\ml{N}[\mf{u}]=1$ and $\ml{M}[\mf{u}]=M$ respectively. 
$\mf{u}$ is at least continuously differentiable. 
\end{thm}

\setcounter{equation}{2}

\begin{cor}\label{cor:porz}
Suppose $\mf u\in\mb{G}_{M,q}$. Then for each $j$, either 
$u_j\equiv 0$ or $u_j>0$ on all of $\mb{R}^3$. 
\end{cor}
\begin{proof}
By \eqref{2.2a}, 
\begin{align}\label{22a2}
\Da u_1 + a_1(x) u_1 = -2 c_s u_0^2 u_{-1}.
\end{align}
where 
\begin{align*}
a_1(x) = \mu +\la -V(x)-2c_n|\mf u|^2-2c_s u_0^2 -u_1^2+u_{-1}^2-q.
\end{align*}
Since $\mf u$ is continuously differentiable and $V(x)$ is locally bounded
(Assumption \ref{A1}), $a_1(x)$ is locally bounded. 
Suppose $u_1(x^*) = 0$ for some $x^*$. 
By subtracting $bu_1$ from both 
sides of \eqref{22a2}, where $b$ is some large enough constant,
we will obtain 
\begin{align*}
\Da u_1 + h(x) u_1 \le 0\,\mbox{ in a domain containing }\,x^*,
\end{align*}
where $h(x)\le 0$ and is locally bounded. 
Note that $u_1(x^*)$ is a nonpositive minimum of $u_1$.
By a standard maximum principle argument
(see e.g. Theorem 6 in p. 64 of \cite{PW84B} or Theorem 8.19 of \cite{Gil_Tru}), 
$u_1\equiv 0$. 
The proofs for $u_0$ and $u_{-1}$ are similar and omitted.
\end{proof}

\begin{cor}\label{cor:max}
Suppose $0<M<1$ and $\mf{u}\in\mb{G}_{M,q}$ is a ground state. 
Then $u_j\ne 0$ (and hence $>0$ everywhere) for $j=1,-1$.
\end{cor}
\begin{proof}
Since $\int(u_1^2-u_{-1}^2)\,dx=M>0$, $u_1\not\equiv 0$, 
and hence $u_1>0$ everywhere
by Corollary \ref{cor:porz}. 
To prove $u_{-1}\ne 0$, assume otherwise, then \eqref{2.2c}
gives $u_0^2 u_1=0$, and so $u_0\equiv 0$. Thus, among the three 
components only $u_1\ne 0$, which implies $M=1$ from the 
constraint $\ml{N}[\mf{u}]=1$, contradicting our assumption.
\end{proof}

In many aspects, our three-component system can be regarded 
as a generalization of the one-component BEC system studied in 
\cite{Lieb-2000}. In particular, all the assertions 
in Theorem \ref{thm:exist} 
can be proved in similar ways to the corresponding results given in 
Appendix A of \cite{Lieb-2000}. Thus, we shall not give the proof of 
the whole theorem, except for the existence result. In fact,
for our purpose, some further
observations from the proof of the existence result will be needed.
Precisely, we will need the following
lemma (which contains the existence result as a corollary). 

\begin{lem}\label{exist_str}
Let $\{\mf{u}^n=(u_1^n,u_0^n,u_{-1}^n)\}_{n\in\mb{N}}$ be a sequence in 
$\mb{B}$, 
$u_j^n\ge 0$ for all $j$ and $n$.
Suppose $\ml{N}[\mf{u}^n]\to 1$, $\ml{M}[\mf{u}^n]\to M$, 
and $\ml{E}[\mf{u}^n]$ is a bounded sequence, then 
$\{\mf{u}^n\}$ has a subsequence 
$\{\mf{u}^{n(k)}\}_{k\in\mb{N}}$ converging weakly in 
$\mb{B}$ to some $\mf{u}^{\infty}\in\mb{A}_M$, which satisfies
\begin{align*}
\ml{E}[\mf{u}^{\infty}]\le
\liminf_{k\to\infty}\ml{E}[\mf{u}^{n(k)}].
\end{align*}
If we assume further 
that $\ml{E}[\mf{u}^n]$ converges to the ground-state energy $E_g$,
then $\mf{u}^{\infty}$ is a ground state, and 
$\mf{u}^{n(k)}\to\mf{u}^{\infty}$ in the norm of $\mb{B}$.
\end{lem} 

What is most special about this lemma is the last assertion of 
strong convergence, which holds for our model due to the fact that each 
part (namely $\ml E_{kin},\ml E_{pot},...$) of the energy functional is nonnegative.
We give the proof of Lemma \ref{exist_str} in the appendix.

\begin{prop}\label{lem:egconti}
For fixed $0\le M\le 1$, $E_g(M,\cdot)$ is a continuous function.
\end{prop}

\begin{proof}
Given any $q_1, q_2\in[0,\infty)$. Let 
$\mf{u}^k=(u_1^k,u_0^k,u_{-1}^k)\in\mb{G}_{M,q_k}$, $k=1,2$. 
Since $\ml{E}[\mf{u}^1,q_1]=E_g(M,q_1)$ and 
$\ml{E}[\mf{u}^1,q_2]\ge E_g(M,q_2)$, we have
\begin{align}\label{lbd}
\begin{aligned}
(q_1-q_2)\int\big[(u_1^1)^2+(u_{-1}^1)^2\big]\,d x
&=\ml{E}[\mf{u}^1,q_1]-\ml{E}[\mf{u}^1,q_2]\\
&\le E_g(M,q_1)-E_g(M,q_2).
\end{aligned}
\end{align}
Similarly
\begin{align}\label{ubd}
\begin{aligned}
E_g(M,q_1)-E_g(M,q_2) 
&\le \ml{E}[\mf{u}^2,q_1]-\ml{E}[\mf{u}^2,q_2]\\
&= (q_1-q_2)\int\big[(u_1^2)^2+(u_{-1}^2)^2\big]\,d x.
\end{aligned}
\end{align}
From (\ref{lbd}) and (\ref{ubd}), we obtain 
\begin{align*}
|E_g(M,q_1)-E_g(M,q_2)| \le |q_1-q_2|,
\end{align*}
and hence the lemma.
\end{proof}

\begin{remark}
$E_g(M,q)$ is also continuous in $M$ \cite{Lin_thesis}, but we will not need this fact. 
\end{remark}

\subsection{The two-component ground state}\label{2cg}

For our three-component system, we do not know if 
ground states are unique in general.
Nevertheless, if we 
consider only the two component case, i.e. $u_0\equiv 0$,
then uniqueness can be proved by a standard 
convexity argument. Precisely, define the two-component admissible class
\begin{align*}
\mb{A}^{two}_M = \lt\{
\mf{u}\in\mb{A}_M\,|\, u_0\equiv 0
\rt\},
\end{align*}
then we have the following uniqueness result. 

\begin{thm}\label{thm:unitwo}
There exists exactly one element in 
$\mb{A}^{two}_M$ which minimizes the 
energy $\ml{E}$ over $\mb{A}^{two}_M$. Moreover, this unique 
minimizer is unchanged for different values of $q$.
\end{thm}

\begin{proof}
The existence assertion can be proved in the same way as 
for the three-component case. To prove the uniqueness result, we exploit 
a standard convexity argument. 
Given $\mf{u},\mf{v}\in\mb{A}^{two}_M$. Let 
$\mf{w}\in\mb{B}$ be defined by 
$w_j^2 = (u_{j}^2+v_{j}^2)/2$ for $j=1,0,-1$, then
$\mf{w}$ also lies in $\mb{A}^{two}_M$. Let 
$D=(\ml{E}[\mf{u}]+\ml{E}[\mf{v}])/2-\ml{E}[\mf{w}]$, then 
$D = D_{kin} + D_n + D_s$, where $D_{kin}$, $D_n$ and $D_s$ 
are the portions corresponding to $\ml{E}_{kin}$, $\ml{E}_n$ and $\ml{E}_s$ 
respectively. We have
\begin{align*}
D_{kin} 
=\int \sum_{j=1,-1}\lt(
\fc{|\na u_j|^2+|\na v_j|^2}{2}-|\na w_j|^2\rt)\,d x\ge 0.
\end{align*}
(For validity of this convexity inequality for gradients, 
see e.g. Theorem 7.8 of \cite{LiebLoss}. Alternatively, it is a corollary of the 
redistribution inequality \eqref{eq:redisineq} in the next section.)
Also, by direct computation we obtain
\begin{align*}
D_n 
=\fc{c_n }{4}\int \big(|\mf{u}|^2-|\mf{v}|^2\big)^2\,d x\ge 0,
\end{align*}
and 
\begin{align*}
D_s
=\fc{c_s}{4}\int \big(
u_1^2-u_{-1}^2-v_1^2+v_{-1}^2
\big)^2\,d x\ge 0.
\end{align*}
Now assume $\mf{u}$ and $\mf{v}$ are both minimizers of 
$\ml{E}$ over $\mb{A}^{two}_M$, then $D\le 0$, and we must 
have $D_{kin}=D_n=D_s=0$. From $D_n=D_s=0$ we conclude that 
$\mf{u} = \mf{v}$, and the uniqueness is proved. 
Finally, as the constraint $\ml{N}[\mf u]=1$ reduces to 
$\int(u_1^2+u_{-1}^2)\,d x = 1$, $\ml{E}_{Zee}$ equals the constant $q$ on
$\mb{A}^{two}_M$. Hence $q$ has no influence on the minimizer.
\end{proof}

\noindent{\bf Notation}.
We will use $\mf z=(z_1,0,z_{-1})$ to denote the unique element
in $\mb{A}_M^{two}$ asserted in Theorem \ref{thm:unitwo}. 

\begin{remarks}~
\begin{enumerate}
\item Although $\mf z$ is independent of $q$, it depends on $M$. For clarity, 
we may use a symbol such as $\mf{z}^M$. We shall however not do so to 
simplify notation.
\item By definition $\mf z$ is the ``ground state'' over $\mb A^{two}_M$,
and is not necessarily a true ground state over $\mb A_M$.
Nevertheless, it is proved in \cite{LinChern2011} that
$\mf{z}$ is indeed a ground state if $q=0$; is the unique ground state
if moreover $0<M\le 1$. 
\end{enumerate}
\end{remarks}

%
%
%
%
%
%

\section{Mass Redistribution and Some Further Properties}\label{sfpro}

In this section, we prove some more facts about ground states.
We will use the method of mass redistribution introduced in 
\cite{LinChern2011}. For convenience, we recap the idea below.

\subsection{Mass redistribution}

Let $\Oa$ be an open set in $\mb{R}^d$, and 
$f_1, f_2,...,f_n, g_1, g_2,..., g_m$ be nonnegative functions in $H^1(\Oa)$
($d$, $n$ and $m$ are arbitrary positive integers).
Then we 
say $\mf{g}=(g_1,...,g_m)$ is a \emph{square redistribution} 
(\emph{redistribution} for short) of $\mf{f}=(f_1,...,f_n)$ 
if 
\begin{align*}
g_i^2 = \sum_{j=1}^n a_{ij}f_j^2\quad (i=1,\ldots,m),
\end{align*}
where the coefficients $a_{ij}$ are nonnegative constants 
satisfying $\sum_{i=1}^m a_{ij} = 1$, for $j=1,...,n$. 
In such a situation, it's obvious that 
$|\mf{g}|=|\mf{f}|$.
Moreover, it is proved that the following pointwise inequality
holds:
\begin{align}\label{eq:redisineq}
\sum_{i=1}^m|\na g_i|^2 \le \sum_{j=1}^n|\na f_j|^2.
\end{align}

We will apply the redistribution technique to our
admissible triples.
More precisely, for a given $\mf u = (u_1,u_0,u_{-1})\in\mb A_M$,
we will consider its redistributions $\mf v=(v_1,v_0,v_{-1})$.
As the squares $u_j^2$ and $v_j^2$ represent mass densities
of the corresponding components, $\mf v$ can be regarded as a 
redistribution of the masses among the three components of $\mf u$.
From the equality $|\mf v|=|\mf u|$ and the inequality \eqref{eq:redisineq}, we have
\begin{align}\label{eq:eqpotn}
\ml{E}_{kin}[\mf{v}]\le\ml{E}_{kin}[\mf{u}],\quad
\ml{E}_{pot}[\mf{v}]=\ml{E}_{pot}[\mf{u}],\quad \mbox{and}\quad
\ml{E}_{n}[\mf{v}]=\ml{E}_{n}[\mf{u}].
\end{align}
Also by $|\mf v|=|\mf u|$, the first constraint $\ml N[\mf v]=1$
is satisfied automatically, and in general we have $\mf v\in\mb{A}_{M'}$ for some 
$M'\in [-1,1]$. 
One particular simple and useful observation is the following fact.

\begin{lem}\label{lem:masred}
If $\mf u\in\mb{G}_{M,q}$, 
and $\mf v$ is a redistribution of $\mf u$ that also lies in the 
same admissible class $\mb A_M$, then
\begin{align*}
\ml{E}_s[\mf{u}]+\ml{E}_{Zee}[\mf{u}]\le
\ml{E}_s[\mf{v}]+\ml{E}_{Zee}[\mf{v}].
\end{align*} 
\end{lem}
\begin{proof}
By assumption, $\ml{E}[\mf{u}]\le \ml{E}[\mf{v}]$, and the assertion is 
a direct consequence of \eqref{eq:eqpotn}.
\end{proof}

\subsection{Pointwise comparisons between $u_1$ and $u_{-1}$}

This section concerns the fact that $u_1$ is larger than $u_{-1}$
everywhere. 
Recall that $E_g(M,q)$ is the ground state energy for given $M,q$,
and $\mb{G}_{M,q}$ is the set of ground state(s). 
We first give a lemma.

\begin{lem}\label{lem:unibdd}
For fixed $q\in[0,\infty)$, $E_g(\cdot,q)$ is a 
strictly increasing function on $[0,1]$.
\end{lem}
\begin{proof}
We first consider $0<M\le 1$. Let $\mf u\in\mb{G}_{M,q}$.
For $0<\da<1$, let $\mf{u}(\da)$ be the redistribution of 
$\mf{u}$ defined by
\begin{align*}
\lt\{
\begin{aligned}
u_1(\da)^2 &= (1-\da)u_1^2\\
u_0(\da)^2 &= \da u_1^2 + u_0^2 + \da u_{-1}^2\\
u_{-1}(\da)^2 &= (1-\da)u_{-1}^2\,.
\end{aligned}
\rt.
\end{align*}
Then $\mf{u}(\da)\in\mb{A}_{(1-\da)M}$.
Since $\mf{u}(\da)$ is a redistribution of $\mf{u}$, we have
$\ml{E}_{kin}[\mf{u}(\da)]\le\ml{E}_{kin}[\mf{u}]$,
$\ml{E}_{pot}[\mf{u}(\da)]=\ml{E}_{pot}[\mf{u}]$, and
$\ml{E}_{n}[\mf{u}(\da)]\le\ml{E}_{n}[\mf{u}]$.
One can also check by direct computation that 
\begin{align*}
\ml{E}_{Zee}[\mf{u}]-\ml{E}_{Zee}[\mf{u}(\da)] 
= q\da\int(u_1^2+u_{-1}^2)\,d x\ge 0,
\end{align*}
and
\begin{align}\label{eq:dofes}
\ml{E}_s[\mf{u}]-\ml{E}_s[\mf{u}(\da)]
=c_s\da\int (u_1-u_{-1})^2\big[
2u_0^2+4u_1u_{-1}+\da(u_1-u_{-1})^2
\big]\,d x> 0.
\end{align}
The strict inequality holds for \eqref{eq:dofes}
since we assume $M>0$ and $u_1-u_{-1}$ cannot be identically zero.
Thus, we obtain
\begin{align*}
E_g((1-\da)M,q)\le\ml{E}[\mf{u}(\da)]<\ml{E}[\mf{u}]=E_g(M,q).
\end{align*}
This is true for all small $\da>0$, and hence $E_g(\cdot,q)$ is 
strictly increasing on $(0,1]$.

To prove that $E_g(\cdot,q)$ is also strictly increasing at $0$,
we first note that $E_g(\cdot,q)$ is a bounded function. 
To prove it, choose any nonnegative function 
$f\in H^1(\mb{R}^3)\cap L^2_V(\mb{R}^3)\cap L^4(\mb{R}^3)$ that satisfies 
$\int f^2 \,d x= 1$, and define
\begin{align*}
\mf{f}^M = ((\fc{1+M}{2})^{1/2}f,0,(\fc{1-M}{2})^{1/2}f).
\end{align*}
We have $\mf{f}^M\in\mb{A}_M$, and hence 
\begin{align*}
E_g(M,q)\le \ml{E}[\mf{f}^M]=\int\Big\{
|\na f|^2 + V f^2 + c_n  f^4 + c_s M^2 f^4 + q
\Big\}\,d x\le \ml{E}[\mf{f}^1],
\end{align*} 
where the upper bound is independent of $M$. 
Now, 
let $\{M_n\}$ be a sequence in $(0,1)$ such that 
$M_n\to 0^+$, and let $\mf{u}^n\in\mb{G}_{M_n,q}$.
Since $\ml{E}[\mf{u}^n]$ is a bounded sequence, by Lemma \ref{exist_str}, there exists
a subsequence $\{\mf{u}^{n(k)}\}$ such that 
$\mf{u}^{n(k)}\wto\mf{u}^{\infty}$ weakly in $\mb{B}$ for 
some $\mf{u}^{\infty}\in\mb{A}_0$. Moreover, 
\begin{align*}
E_g(0,q) \le \ml{E}[\mf{u}^{\infty}] 
         \le \liminf_{k\to\infty}\ml{E}[\mf{u}^{n(k)}]
           = \liminf_{k\to\infty}E_g(M_{n(k)},q)
           = \inf_{0< M\le 1}E_g(M,q).
\end{align*}
The last equality is due to the monotonicity of 
$E_g(\cdot,q)$ on $(0,1]$. Thus $E_g(0,q)< E_g(M,q)$ for 
every $M\in(0,1]$. Here the strict inequality holds since otherwise
we would obtain
$E_g(0,q)=E_g(M,q)>E_g(M/2,q)\ge E_g(0,q)$, a contradiction.
\end{proof}

\begin{prop}\label{prop:gtreq}
For every $0\le M\le 1$ and $q\ge 0$, $\mf{u}\in\mb{G}_{M,q}$ 
satisfies $u_{-1}\le u_1$.
\end{prop}

\begin{proof}
Let $\mf{v}=(v_1,v_0,v_{-1})$ be defined by $v_1 = \max(u_1,u_{-1})$, 
$v_{-1} = \min(u_1,u_{-1})$, and $v_0=u_0$. 
Then 
$\ml{E}[\mf{v}]=\ml{E}[\mf{u}]$. The only nontrivial part for this assertion is 
the equality for $\ml{E}_{kin}$. To verify it, one can use the 
formula 
\begin{align*}
v_j = \fc{1}{2}\lt(u_j+u_{-j}+j|u_j-u_{-j}|\rt),
\quad\mbox{for}\quad j=1,-1.
\end{align*}
Then 
\begin{align}\label{eqofkin}
|\na v_1|^2+|\na v_{-1}|^2
=\fc{1}{2}\big(
|\na u_1|^2 + |\na u_{-1}|^2 + 2\na u_1\cdot\na u_{-1}
+2\big|\na|u_1-u_{-1}|\big|^2
\big).
\end{align}
Since $|\na |f||=|\na f|$ a.e. 
for general real-valued $W^{1,p}$ functions $f$ (see e.g. Theorem 6.17 of \cite{LiebLoss}),
\begin{align*}
\big|\na|u_1-u_{-1}|\big|^2 = |\na u_1|^2 - 2\na u_1\cdot\na u_{-1}+ |\na u_{-1}|^2.
\end{align*}
Taking this into \eqref{eqofkin}, we obtain 
$|\na v_1|^2+|\na v_{-1}|^2=|\na u_1|^2+|\na u_{-1}|^2$, and
hence $\ml{E}_{kin}[\mf{v}]=\ml{E}_{kin}[\mf{u}]$.

Thus, we have
\begin{align*}
E_g(\ml{M}[\mf{v}],q)\le\ml{E}[\mf{v}]=\ml{E}[\mf{u}]=E_g(M,q).
\end{align*}
As $E_g(\cdot,q)$ is strictly increasing, the above relation 
implies 
\begin{align*}
\int (v_1^2-v_{-1}^2)\, dx = \ml{M}[\mf{v}]\le M = \int (u_1^2-u_{-1}^2)\, d x.
\end{align*}
By definition of $\mf v$, this is possible only if 
$u_1^2-u_{-1}^2 = v_1^2-v_{-1}^2\ge 0$, and the proof is completed.
\end{proof}

Proposition \ref{prop:stronggtr} and Proposition \ref{prop:mzeroqpo} below
give more precise claims than Proposition \ref{prop:gtreq} in different situations. 

\begin{lem}\label{lanonneg}
If $0<M<1$ and $\mf{u}\in\mb{G}_{M,q}$, the Lagrange 
multiplier $\la$ in \up{(\hyperref[2.2a]{2.2})}
is positive.
\end{lem}
\begin{proof}
\eqref{2.2a} multiplied by $u_{-1}$ minus \eqref{2.2c} 
multiplied by $u_1$ gives
\begin{align*}
2\la u_1u_{-1} = \na\cdot(-u_{-1}\na u_1+u_1\na u_{-1}) 
+ 2c_s(u_1^2-u_{-1}^2)(u_0^2+2u_1u_{-1}).
\end{align*}
Taking integration, we get 
\begin{align}\label{intlaeq}
\la\int u_1u_{-1} \,d x
= c_s\int(u_1^2-u_{-1}^2)(u_0^2+2u_1u_{-1})\, dx.
\end{align}
By Corollary \ref{cor:max}, $u_1u_{-1}>0$. On the other 
hand, since $M>0$, $u_1^2-u_{-1}^2$ can not be identically zero. 
Hence \eqref{intlaeq} implies $\la>0$.
\end{proof}

\begin{prop}\label{prop:stronggtr}
If $0<M\le 1$ and $q\ge 0$, then $\mf{u}\in\mb{G}_{M,q}$ 
satisfies $u_{-1} < u_1$. 
\end{prop}

\begin{proof}
Let $w=u_1 - u_{-1}$. Then \eqref{2.2a} minus \eqref{2.2c} gives
\begin{align}\label{weq}
\Da w + Q w = -\la(u_1+u_{-1}) - \mu w,
\end{align}
where
\begin{align*}
Q = -V-2c_n |\mf{u}|^2-2c_s\lt[
2u_0^2+(u_1+u_{-1})^2\rt]-q \le 0.
\end{align*}
Since $\la> 0$ and $w\ge 0$, by subtracting $|\mu|w$ 
from both sides of (\ref{weq}), we obtain 
\begin{align*}
\Da w + \wt{Q}w \le 0,
\quad\mbox{where}\quad \wt{Q}=Q-|\mu|\le 0.
\end{align*}
By the strong maximum principle, either $w>0$ everywhere or 
$w\equiv 0$. Since $M>0$, $w>0$.
\end{proof}

\begin{prop}\label{prop:mzeroqpo} 
For any $q\ge 0$ and $\mf u\in\mb{G}_{0,q}$ (i.e. $M=0$),  
$u_1=u_{-1}$. If moreover $q>0$, then $u_1=u_{-1}\equiv 0$. 
\end{prop}
\begin{proof}
The equality $u_1=u_{-1}$ follows from $u_{-1}\le u_1$ and the assumption $M=0$.
To see why they must vanish when $q>0$, consider 
the redistribution $(0,|\mf u|, 0)$ of $\mf u$. Note that $(0,|\mf u|,0)\in\mb A_0$.
By Lemma \ref{lem:masred} and 
the fact $E_s[\mf u]=E_s[(0,|\mf u|,0)]=\ml E_{Zee}[(0,|\mf u|,0)]=0$,
we obtain 
\begin{align*}
\ml E_{Zee} [\mf u] =q\int (u_1^2+u_{-1}^2)\,d x = 0.
\end{align*}
Hence $u_1=u_{-1}=0$.
\end{proof}

\begin{remark}
\item In the above proof we obtained $u_1=u_{-1}$ from 
the knowledge of $u_{-1}\le u_1$. 
Another simple and direct method is to consider the redistribution 
$\mf v$ defined by $v_1^2 = v_{-1}^2 = (u_1^2+u_{-1}^2)/2$,
and $v_0^2 = u_0^2$. 
Then $\ml E_s[\mf v]=0$ and $\ml E_{Zee}[\mf v] = \ml E_{Zee}[\mf u]$. 
By Lemma \ref{lem:masred}, we get $\ml E_s[\mf u]\le 0$, and 
the assertion follows.
\end{remark}

\begin{remark}\label{rem:mzero}
Similar to the proof of Theorem \ref{thm:unitwo}, we can prove that there is a unique 
minimizer of $\ml E$ over the one-component class 
$\{\mf u\in\mb{A}_0\,|\, u_1=u_{-1}\equiv 0\}$. Thus, Proposition \ref{prop:mzeroqpo} implies that $\mf{G}_{0,q}$ has a unique element for $q>0$.
By contrast, at $M=q=0$, it is shown in \cite{LinChern2011} (Proposition 4.2) that
ground states are not unique, and $u_1, u_{-1}$ may be nonzero.
Also note that at the other endpoint $M=1$, we have $u_0=u_{-1}\equiv 0$,
and it can also be proved that there exists a unique one-component ground state
of the form $(u_1,0,0)$. 
\end{remark}

%
%
%
%
%
%

\section{The Phase Transition Phenomenon}\label{sec:ptp}

\subsection{Proof of the Main Theorem}

We prove Theorem \ref{thm:qc} in this section. 
Let $0<M<1$ be fixed. 
Recall from Section \ref{2cg} that 
$\mb{A}^{two}_M$ is the 2C admissible class, and 
$\mf z=(z_1,0,z_{-1})$ denotes the unique minimizer of $\ml E$ over 
$\mb{A}^{two}_M$. 

We will prove the theorem by defining
\begin{align}\label{defqc}
q_c(M) = \sup\lt\{q \,\lt|\, \mf{z}\in\mb{G}_{M,q}\rt.\rt\}.
\end{align}
The theorem is divided into four claims.

\begin{claim}\label{clm1}
$q_c(M)<\infty$.
\end{claim}

\begin{proof}
Suppose $\mf z\in\mb{G}_{M,q}$. The idea is to get a restriction on $q$ 
by Lemma \ref{lem:masred}. For example,
consider the redistribution
$\mf v$ of $\mf z$ defined by 
\begin{align*}
\lt\{\begin{aligned}
v_1^2 &= (1-r)z_1^2\\
v_0^2 &= r z_1^2 + z_{-1}^2\\
v_{-1}^2 &= 0\,,
\end{aligned}\rt.
\end{align*}
where $r=(1-M)/(1+M)$. Then $\mf{v}\in\mb{A}_M$, and Lemma \ref{lem:masred}
implies
\begin{align}\label{upi}
&c_s\int(z_1^2-z_{-1}^2)^2\,dx + q \\
&\qquad \le 
c_s\int\left[2(rz_1^2+z_{-1}^2)(1-r)z_1^2+(1-r)^2z_1^4\right]\,dx
+q\int(1-r)z_1^2\,dx.
\end{align}
Since $\mf z$ is independent of $q$, the above inequality gives an upper bound on $q$.
\end{proof}

\begin{remark}\label{acm1}
Note that the constraints 
$\ml N[\mf z]=1$ and $\ml M[\mf z]=M$ are equivalent to 
\begin{align*}
\int z_1^2\,dx = (1+M)/2,\quad \int z_{-1}^2\,dx = (1-M)/2,
\end{align*}
from which \eqref{upi} can be arranged into the 
more clear form
\begin{align}\label{rindofq}
q\le \fc{c_s}{1-M} \int\lt\{
-r^2 z_1^4 + (4-2r)z_1^2z_{-1}^2-z_{-1}^4
\rt\}\,dx.
\end{align}
The right-hand side of \eqref{rindofq} is an upper bound on $q$ for $\mf{z}$ to be a ground 
state, and hence is an upper bound for $q_c(M)$. 
\end{remark}

\begin{claim}\label{clm2}
$q_c(M)>0$.
\end{claim}
The proof of this claim is the most difficult part of the whole proof, and we give it 
in an independent section below. 

\begin{claim}\label{clm3}
Let $q>q_c(M)$ and $\mf u\in\mb{G}_{M,q}$. Then $\mf u$ is a 3C state.
\end{claim}
\begin{proof}
This is a direct consequence from 
our definition \eqref{defqc} of $q_c(M)$ and Corollary \ref{cor:max}.
\end{proof}

Finally, it remains to show the following. 

\begin{claim}\label{clm4}
Let $q<q_c(M)$ and $\mf u\in\mb{G}_{M,q}$. Then $\mf u=\mf z$.
\end{claim}

\begin{proof}
Let us here write $\ml{E}[\mf{u},q]$ to indicate the value of $q$.
By definition of $q_c(M)$, there 
exists $q<q'<q_c(M)$ such that $\mf z\in\mb{G}_{M,q'}$.
Thus
\begin{align}\label{thei}
\begin{aligned}
\ml{E}[\mf{z},q']\le 
\ml{E}[\mf{u},q'] 
&= \ml{E}[\mf{u},q]+(q'-q)\int \big(u_1^2+u_{-1}^2\big)\,dx
\\
&\le\ml{E}[\mf{z},q] + (q'-q)\int \big(z_1^2+z_{-1}^2\big)\,dx
=\ml{E}[\mf{z},q'].
\end{aligned}
\end{align}
The inequality is true since $\mf u\in \mb{G}_{M,q}$ and since 
$\int (u_1^2+u_{-1}^2)\,dx\le 1 = \int(z_1^2+z_{-1}^2)\,dx$.
As the leftmost term and the rightmost term in \eqref{thei} are the same,
the inequality is actually an equality.
In particular, $\int (u_1^2+u_{-1}^2)\,dx= 1$, which means $u_0\equiv 0$.
Now that $\mf u\in\mb{G}_{M,q}$ is a 2C state, it must also be a minimizer 
of $\ml E$ over $\mb{A}^{two}_M$, and hence $\mf u=\mf z$.
\end{proof}

\subsubsection{Proof of Claim 2}\label{sec:pc2}

We will need several facts, which we collect in 
a single lemma below. (Recall that $0<M<1$ is fixed.)

\begin{lem}\label{lemab}
Given $q_n\to q_\infty$ in $[0,\infty)$ and 
$\mf{u}^n\in \mb{G}_{M,q_n}\to \mf{u}^\infty\in\mb{G}_{M,q}$
in the norm of $\mb{B}$.
The following assertions are true.
\begin{itemize}
\item[(i)] The Lagrange multipliers $\mu_n$ and $\la_n$ in (\hyperref[2.2a]{2.2})
corresponding to $\mf{u}^n$ converge respectively to 
$\mu_{\infty}$ and $\la_{\infty}$, those corresponding to $\mf{u}^\infty$.
\item[(ii)] For any $\vn>0$,
there exists $r_j>0$ ($j=1,0,-1$) independent of $n$ such that 
$u_j^n(\mf x)\le \vn$ and $u_j^\infty(x)\le \vn$ for $|x|\ge r_j$.
\item[(iii)] There exists $R>0$ independent of $n$ such that 
\begin{align}\label{aslema}
\fc{1}{2}\int (u_0^n)^2\,dx \le \int_{B(R)} (u_0^n)^2\,dx\quad
\up{for all }\,n,
\end{align}
where $B(R)$ denotes the open ball with center the origin and radius $R$. 
\item[(iv)] $\mf{u}^n\to\mf{u}^{\infty}$ uniformly.
\end{itemize}
\end{lem}
The proof of the whole lemma is lengthy, and we postpone it to the 
end of this section. Let us now assume it and prove Claim \ref{clm2}.

Let $\mf{u}=(u_1,u_0,u_{-1})\in\mb{G}_{M,q}$. 
Consider the redistribution $\mf{v}\in\mb{A}_M$ of $\mf{u}$ defined by
\begin{align*}
\lt\{\begin{aligned}
v_1^2 &= u_1^2+\fc{1}{2}u_0^2\\
v_0^2 &= 0\\
v_{-1}^2 &= u_{-1}^2+\fc{1}{2}u_0^2\,.
\end{aligned}\rt.
\end{align*}
Then Lemma \ref{lem:masred} implies
\begin{align}\label{ztheg}
q\int u_0^2\,dx \ge 2c_s\int u_0^2(u_1-u_{-1})^2\,dx.
\end{align}
It is obvious that $u_0\equiv 0$ (and hence $\mf{u}=\mf{z}$) if $q=0$.
This is exactly the argument used in \cite{LinChern2011} to show that 
$\mf z$ is the unique element of $\mb{G}_{M,0}$.
For $q>0$, however, the situation becomes trickier. 
We will apply \eqref{ztheg} to a sequence $\mf{u}^n$
as described in the following claim, and show that 
$u^n_0\equiv 0$ for large $n$.

\begin{claim*}
There exists a sequence $\mf{u}^n=(u_1^n,u_0^n,u_{-1}^n)\in\mb{G}_{M,q_n}$,
where $q_n\to 0$ and $q_n\ne 0$ for all $n$,
such that $\mf{u}^n\to\mf{z}$ in $\mb{B}$.
\end{claim*}
\begin{proof}
Given any sequence $\mf{u}^n\in\mb{G}_{M,q_n}$, where 
$q_n\to 0$ and $q_n\ne 0$ for all $n$.
We have $\ml{N}[\mf{u}^n]= 1$ and $\ml{M}[\mf{u}^n]=M$ for all $n$. 
Moreover, by the continuity of $E_g(M,\cdot)$ (Proposition \ref{lem:egconti}),
\begin{align*}
\ml{E}[\mf{u}^n,0]=E_g(M,q_n)-q_n\int [(u_1^n)^2+(u_{-1}^n)^2]\,dx
\to E_g(M,0)
\end{align*}
Thus, Lemma \ref{exist_str}
implies that $\{\mf{u}^n\}$ has a subsequence $\{\mf{u}^{n(k)}\}_{k\in\mb{N}}$ 
converging to some $\mf{u}^\infty\in\mb{G}_{M,0}$.
However, $\mf z$ is the only element in $\mb{G}_{M,0}$, and hence 
the sequence $\mf{u}^{n(k)}\in\mb{G}_{M,q_{n(k)}}$ 
satisfies the assertion to be proved.
\end{proof}

Now let $q_n\to q_\infty\coloneqq 0$ and
$\mf{u}^n\in\mb{G}_{M,q_n}\to\mf{u}^\infty\coloneqq\mf{z}\in\mb{G}_{M,0}$ 
be as in the claim above. 
For this sequence, let $R$ be the corresponding radius 
asserted in Lemma \ref{lemab} (iii), and let 
$a = \inf_{B(R)}(z_1-z_{-1})$. 
By Proposition \ref{prop:stronggtr}, $a>0$.
By Lemma \ref{lemab} (iv), $\mf{u}^n\to \mf{z}$ uniformly, 
and hence $(u_1^n-u_{-1}^n)\ge a/2$ on $B(R)$ for large $n$.
Thus, we obtain
\begin{align}\label{aaeq}
\begin{aligned}
\int (u_0^n)^2(u_1^n-u_{-1}^n)^2 \,dx
&\ge \int_{B(R)} (u_0^n)^2(u_1^n-u_{-1}^n)^2\,dx\\ 
&\ge \fc{a^2}{4}\int_{B(R)} (u_0^n)^2\,dx
\ge \fc{a^2}{8}\int (u_0^n)^2\,dx
\end{aligned}
\end{align}
for large $n$. On the other hand, (\ref{ztheg}) implies 
\begin{align}\label{bbeq}
q_n\int (u_0^n)^2\,dx\ge 2c_s\int (u_0^n)^2(u_1^n-u_{-1}^n)^2\,dx
\end{align}
for all $n$. Since $q_n\to 0$, (\ref{aaeq}) and (\ref{bbeq}) 
together imply that
$u_0^n\equiv 0$, and the proof of Claim \ref{clm2} is completed.

Finally, we prove Lemma \ref{lemab}.

\begin{proof}[Proof of Lemma \ref{lemab}]\ 

[Proof of (i)]
Multiplying \eqref{2.2a} by $u_1$ and multiplying \eqref{2.2c} by $u_{-1}$, 
and then taking integration, we obtain 
\begin{align}\label{tosol}
(\mu+j\la)\int u_j^2\,dx = F_j(\mf{u},q) \quad (j=1,-1),
\end{align}
where
\begin{align*}
F_j(\mf{u},q) 
&= \int \Big\{|\na u_j|^2+V(x)u_j^2 + 2c_n  |\mf{u}|^2 u_j^2
\\
&\qquad +2c_s\big[u_0^2u_j(u_j-u_{-j})+u_j^2(u_j^2-u_{-j}^2)\big]
+qu_j^2\Big\}\,dx.
\end{align*}
We can solve (\ref{tosol}) for $\mu$ and $\la$ as long as 
$\int u_1^2\,dx$ and $\int u_{-1}^2\,dx$ are positive, and obtain
\begin{align}\label{explicit}
\begin{aligned}
\mu &= \textstyle{
\big[F_1(\mf{u},q)/(\int u_1^2\,dx)+F_{-1}(\mf{u},q)/\int u_{-1}^2\,dx\big]/2}\\
\la &= \textstyle{
\big[F_1(\mf{u},q)/(\int u_1^2\,dx)-F_{-1}(\mf{u},q)/\int u_{-1}^2\,dx\big]/2}.
\end{aligned}
\end{align}
By Corollary \ref{cor:max}, $u_j^n$ and $u_j^\infty$ ($j=1,-1$) are positive functions.
Thus (\ref{explicit}) applies to
$\mu_n,\la_n$ and $\mu_{\infty},\la_{\infty}$, and 
$\mu_n\to\mu_\infty$ and $\la_n\to\la_\infty$ follow from the 
convergence $\mf{u}^n\to\mf{u}^\infty$ in $\mb{B}$.

[Proof of (ii)] (For this proof we consider $n\in\mb{N}\cup\{\infty\}$.)
By \eqref{2.2b}, $\Da u_0^n = f_0^n$, 
where 
\begin{align*}
f_0^n = [-\mu^n + V(x) +2c_n |\mf{u}^n|^2 + 2c_s(u_1^n-u_{-1}^n)^2]u_0^n.
\end{align*}
By (i), $\{\mu^n\}$ is bounded, and Assumption \ref{A1} ensures that
there is an $R_0>0$ independent of $n$ such that $\Da u_0^n(x) \ge 0$ 
(subharmonicity) for $|x|\ge R_0$. Thus for $|x|>R_0$, we have the mean 
value inequality
\begin{align*}
u_0^n(x) \le |B(x,|x|-R_0)|^{-1}\int_{B(x, |x|-R_0)} u_0^n(y)dy.
\end{align*}
Here $B(x,\rho)$ denotes the open ball with center $x$ and radius $\rho$, 
and $|B(x,\rho)|$ denotes its volume. Since 
$\|u_0^n\|_{L^2}\le \|\mf{u}^n\|_{L^2}=1$, by H\"{o}lder's inequality, 
we get 
\begin{align*}
u_0^n(x) \le |B(x, |x|-R_0)|^{-1/2}\|u_0^n\|_{L^2}
\le |B(0,1)|^{-1/2}(|x|-R_0)^{-3/2}.
\end{align*}
Thus $u_0^n(x)\le \vn$ for 
$|x|\ge r_0\coloneqq (|B(0,1)|^{1/2}\vn)^{-2/3}+R_0$.

For $u_1^n$, from the fact $u_{-1}^n\le u_1^n$, we see \eqref{2.2a} also implies
$\Da u_1^n(x) \ge 0$ for $|x|$ larger than some $R_1>0$. Thus 
similarly we can show that $u_1^n(x)\le \vn$
for $|x|\ge r_1\coloneqq (|B(0,1)|^{1/2}\vn)^{-2/3}+R_1$. In 
contrast, the subharmonicity of $u_{-1}^n$ outside some ball is not so obvious 
from \eqref{2.2c}. 
Nevertheless, since $u_{-1}^n\le u_1^n$, we may just choose $r_{-1}=r_1$.

[Proof of (iii)]
By (i), $\mu_n\le C$ for some constant $C>0$.
Multiplying \eqref{2.2b} for $\mf{u}^n$ by $u_0^n$, 
and taking integration, we obtain
\begin{align*}
\mu_n\int (u_0^n)^2\,dx &= \int\Big\{
|\na u_0^n|^2 + V(x)(u_0^n)^2\\ 
&\qquad\qquad + 2c_n |\mf{u}^n|^2(u_0^n)^2 
+ 2c_s (u_0^n)^2(u_1^n-u_{-1}^n)^2 
\Big\}\,dx,
\end{align*}
which implies
\begin{align}\label{eq1}
\int V(x)(u_0^n)^2\,dx \le \mu_n\int (u_0^n)^2\,dx \le C\int (u_0^n)^2\, dx.
\end{align}
On the other hand, by the assumption \ref{A1}, there exists $R>0$ 
such that $V(x)\ge 2C$ for $|x|\ge R$, and hence
\begin{align}\label{eq2}
\int V(x)(u_0^n)^2\, dx \ge \int_{|x|\ge R}V(x)(u_0^n)^2 \,dx
\ge 2C\int_{|x|\ge R}(u_0^n)^2\,dx.
\end{align}
From (\ref{eq1}) and (\ref{eq2}), we obtain 
\begin{align*}
\int (u_0^n)^2\,dx \ge 2\int_{|x|\ge R}(u_0^n)^2\,dx 
= 2\bigg(\int (u_0^n)^2\,dx - \int_{B(R)} (u_0^n)^2\,dx\bigg),
\end{align*}
which implies the assertion.

[Proof of (iv)]

Let $v_1^n = u_1^n-u_1^{\infty}$. 
Subtract \eqref{2.2a} for $\mf{u}^{\infty}$ from \eqref{2.2a} 
for $\mf{u}^n$, we obtain
\begin{align}\label{dnl}
\Da v_1^n - V(x)v_1^n = P_n -P_{\infty} + S_n - S_{\infty},
\end{align}
where
\begin{align*}
P_n &= -(\mu_n+\la_n-q_n)u_1^n,\\
S_n &= 2c_n  |\mf{u}^n|^2 u_1^n + 2c_s \big[
(u_0^n)^2(u_1^n-u_{-1}^n) + u_1^n\big((u_1^n)^2-(u_{-1}^n)^2\big)
\big],
\end{align*}
and $P_{\infty}$ and $S_{\infty}$ are given by the same expressions 
with $n$ replaced by $\infty$ ($q_{\infty}$ is understood to be $q$).
By the global boundedness result for elliptic operators
(see e.g. Theorem 8.16 of \cite{Gil_Tru}), (\ref{dnl}) implies that, 
for every $r>0$,
\begin{align}\label{globdd}
\sup_{B(r)} |v_1^n| \le \sup_{\pl B(r)}|v_1^n| 
                        + C\|P_n-P_{\infty}+S_n-S_{\infty}\|_{L^2},
\end{align}
where the constant $C>0$ depends only on the radius $r$ and 
$\sup_{B(r)}V$. Now since $q_n\to q$, $\mu_n\to\mu_{\infty}$, 
$\la_n\to\la_{\infty}$, and $\mf{u}^n\to\mf{u}^{\infty}$ in $\mb{B}$, 
we see $P_n\to P_{\infty}$ and $S_n\to S_{\infty}$ in $L^2$ 
(for the latter, we uses the continuous embedding $H^1\to L^6$). 
On the other hand,
By (ii), given $\vn>0$, there exists $r_1>0$ independent
of $n$ such that 
\begin{align}\label{outer}
\sup_{|x|\ge r_1}|v_1^n(x)|
\le \sup_{|x|\ge r_1}(|u_1^n(x)|+|u_1^\infty(x)|)\le 2\vn.
\end{align}
In particular $\sup_{\pl B(r_1)}|v_1^n|\le 2\vn$.
Thus, by letting $r=r_1$ in (\ref{globdd}), and letting $n\to\infty$, 
we obtain
\begin{align}\label{inner}
\limsup_{n\to\infty} \Big(\sup_{x\in B(r_1)} |v_1^n(x)|\Big) \le 2\vn.
\end{align}
Combining (\ref{outer}) and (\ref{inner}), we get
\begin{align*}
\sup_{x\in\mb{R}^3}|v_1^n(x)|\le 3\vn\quad\up{for }\,n\,\up{ large enough}.
\end{align*}
Since $\vn>0$ is arbitrary, we conclude that $v_1^n\to 0$ uniformly 
on $\mb{R}^3$. Similarly we can prove $v_0^n=u_0^n-u_0^\infty$ and 
$v_{-1}^n=u_{-1}^n-u_{-1}^{\infty}$ converge 
to zero uniformly, and the proof is completed.
\end{proof}

\subsection{Some additional issues}

We discuss some additional issues in this section.

\subsubsection{An upper bound of $q_c(M)$ by redistributional perturbation}

In Remark \ref{acm1} we give an upper bound of $q_c(M)$ in terms of 
$\mf{z}$.
Such an upper bound may be of interest
in view of the fact that 
$\mf z$ is independent of $q$ and that
it can be obtained by minimizing a two-component energy functional
(hence simpler than the original three-component problem).
As a consequence, it would also be of interest to sharpen the 
upper bound. 
In Remark \ref{acm1}, that bound is deduced from the 
redistribution $\mf v$ of $\mf z$
given in the proof of Claim \ref{clm1}.
Here we introduce another approach.
Instead of choosing the specific $\mf v$, consider 
a family of redistributions $\mf{u}(\da)=(u_1(\da),u_0(\da),u_{-1}(\da))$ 
of $\mf{z}$, defined by
\begin{align*}
\lt\{\begin{aligned}
u_1(\da)^2 &= (1-r\da)z_1^2\\
u_0(\da)^2 &= r\da z_1^2 + \da z_{-1}^2\\
u_{-1}(\da)^2 &= (1-\da)z_{-1}^2\,,
\end{aligned}\rt.
\end{align*}
where $r = (1-M)/(1+M)$. Then 
$\mf{u}(\da)\in\mb{A}_M$ for all $0\le\da\le 1$. 
Note that $\mf{u}(0)=\mf z$, and $\mf{u}(1)=\mf v$. 
Thus $\mf{u}(\da)$ is a perturbation of $\mf z$ ``towards $\mf v$''.

Define 
\begin{align*}
I(\da) = \ml{E}_s[\mf{u}(\da)]+\ml{E}_{Zee}[\mf{u}(\da)].
\end{align*}
From Lemma \ref{lem:masred}, $\da=0$ is an 
endpoint minimum of $I$, and hence $I'(0^+)\ge 0$. 
By direct computation, this gives
\begin{align}\label{qcupb2}
q\le 4c_s\int z_1 z_{-1}(z_1 - z_{-1})
\Big(\frac{z_{-1}}{1-M} - \frac{z_1}{1+M}\Big)\,dx.
\end{align}
The right-hand side of \eqref{qcupb2} is hence an upper bound 
of $q_c(M)$.
This approach looks more elaborate, and numerical results show that
the upper bound is indeed better (i.e. smaller) than the one 
given in Remark \ref{acm1}. However, we do not have analytical proof of it. 

\subsubsection{Description of the function $q_c(M)$}

In this paper, we only prove the existence of 
the number $q_c(M)$ (for $0<M<1$), which separates the phase 
of 2C ground states
and that of 3C ground states. Some more properties 
about $q_c(M)$ that can be postulated from numerical simulations
(see e.g. Figure 5 of \cite{Lim2008.12}) 
are left open.
For example, it looks like $q_c(M)$ is a continuous (indeed, smooth to some degree) and
strictly increasing function of $M$, with $q_c(0^+)=0$. 
Nevertheless, let us mention that one fact that is not quite clear from numerical 
simulations is proved in \cite{Lin_thesis}, namely
$\lim_{M\to 1}q_c(M)<\infty$.

\subsubsection{The situation at $q_c(M)$}\label{atqc}

As is mentioned, we do not know whether ground states are unique 
in general. Precisely, we do not know it
for $M\in(0,1)$, $q\ge q_c(M)$.
In particular, 
we do not exclude the possibility that there are, besides 
the 2C ground state $\mf{z}$,
other 3C ground states at $q=q_c(M)$ ($0<M<1$).
To have a better idea of the problem, 
recall (from Remark \ref{rem:mzero}) that there is really a ``nonuniqueness point'': 
$(M,q)=(0,0)$, 
which joints two phases that sharply contrast each other: 
For $0<M\le 1$ and $q=0$, the unique ground state is 
$\mf z=(z_1,0,z_{-1})$, 
while for $M=0$ and $q>0$, the unique ground state is of the form 
$(0,u_0,0)$. 
It is observed in numerical simulations that such 
sharp contrast occurs along the curve $q_c(M)$ ($0<M<1$)
on the $(q,M)$-phase diagram.
In \cite{Lim2008.12}, it is then claimed that there are 
both 2C and 3C ground states at $q=q_c(M)$, and hence ground states are not unique there.
Nevertheless, from some other simulations (see e.g. \cite{chen2015}),
it looks possible to track the ground state from the 2C profile to a 3C profile as 
$q$ passes $q_c(M)$, only that $u_0$ grows out rapidly. In this 
scenario, $\mf z$ is the unique ground state at $q=q_c(M)$. 

%
%
%
%
%
%
%
%

\section*{Appendix: Proof of Lemma \ref{exist_str}}

We first remark that with the norm defined by (\ref{normofB}), 
$\mb{B}$ is a reflexive Banach space, in which weak convergence 
is equivalent to weak convergence in $H^1(\mb R^3)$, 
in $L_V^2(\mb R^3)$, 
and in $L^4(\mb R^3)$ separately. We omit the verification of these standard facts. 
We will use $\mb{B}_+$ to denote the subclass of $\mb{B}$ consisting 
of all $(u_1,u_0,u_{-1})$ with $u_j\ge 0$ for all $j=1,0,-1$. 
Since $\mb{B}_+$ is a convex and closed subset of $\mb{B}$, 
$\mb{B}_+$ is a weakly closed subset of $\mb{B}$ (Mazur's theorem).

Since the five parts $\ml{E}_{kin}$, $\ml{E}_{pot}$, etc. are 
nonnegative functionals on $\mb{B}$, the boundedness of 
$\ml{E}[\mf{u}^n]$ implies that of the five parts, and hence 
$\{\mf{u}^n\}$ is a bounded sequence in $\mb{B}$. By the reflexivity 
of $\mb{B}$, $\{\mf{u}^n\}$ has a weakly convergent subsequence 
$\{\mf{u}^{n(k)}\}_{k\in\mb{N}}$. Denote the weak limit by $\mf{u}^{\infty}$. We have 
$\mf{u}^{\infty}\in\mb{B}_+$ since $\mb{B}_+$ is weakly closed.

We now prove $\mf{u}^{\infty}\in\mb{A}_M$. Since $\{\mf{u}^n\}$ is 
bounded in $\mb B$, $\int V(x)|\mf{u}^{n(k)}|^2\,dx \le C$ for some $C>0$ 
independent of $n$. By the assumption \ref{A1}, for any $\vn>0$, there 
exists $R_{\vn}>0$ such that $V(x)\ge C/\vn$ for $|x|\ge R_{\vn}$. 
Thus we have
\begin{align*}
C\ge \int V(x)|\mf{u}^{n(k)}|^2\,dx \ge \int_{B(R_{\vn})^c}V(x)|\mf{u}^{n(k)}|^2 \,dx
\ge \fc{C}{\vn}\int_{B(R_{\vn})^c}|\mf{u}^{n(k)}|^2\,dx,
\end{align*}
and hence $\int_{B(R_{\vn})^c}|\mf{u}^{n(k)}|^2\,dx\le \vn$ for all $k$. 
The weak convergence of $\mf{u}^{n(k)}$ in 
$H^1(\mb R^3)$ implies that, for each $j=1,0,-1$, $u_j^{n(k)}\wto u_j^{\infty}$ weakly in 
$L^2(\mb{R}^3)$, 
and $u_j^{n(k)}\to u_j^{\infty}$ strongly in $L^2(B(R_{\vn}))$ by the 
compact embedding of $H^1(B(R_{\vn}))$ in $L^2(B(R_{\vn}))$. Hence, 
by the weak lower-semi-continuity of norms, we have
\begin{align}\label{uty}
\begin{aligned}
\int (u^{\infty}_j)^2 \,dx
&\le \liminf_{k\to\infty}\int (u^{n(k)}_j)^2\,dx\\
&\le \limsup_{k\to\infty}\int (u^{n(k)}_j)^2\,dx\\
&= \limsup_{k\to\infty}\Big(
   \int_{B(R_{\vn})^c}(u^{n(k)}_j)^2\,dx
   + \int_{B(R_{\vn})}(u^{n(k)}_j)^2 \,dx
   \Big)\\
&\le \vn + \int_{B(R_{\vn})}(u^{\infty}_j)^2\,dx
 \le \vn + \int(u^{\infty}_j)^2\,dx.
\end{aligned}
\end{align}
Since $\vn>0$ is arbitrary, the first and the second inequalities of (\ref{uty}) must be equalities. Thus, 
\begin{align}\label{cofnorm}
\lim_{k\to\infty}\int(u^{n(k)}_j)^2\,dx = \int(u^{\infty}_j)^2\,dx,
\end{align}
and $\mf{u}^{\infty}\in\mb{A}_M$.

Next, 
since $\mf{u}^{\infty}$ is the weak limit of $\mf{u}^{n(k)}$, 
by the standard lower-semi-continuity result (see e.g. 
Theorem 1.6 of \cite{Struwe2008}), we have
\begin{align*}
\ml{E}[\mf{u}^{\infty}]\le\liminf_{k\to\infty}\ml{E}[\mf{u}^{n(k)}].
\end{align*}
It remains to show the last assertion of strong convergence.
Indeed, also by Theorem 1.6 of \cite{Struwe2008}, we have
\begin{align*}
\int|\na u_j^{\infty}|^2\,dx &\le \liminf_{k\to\infty}\int|\na u_j^{n(k)}|^2\,dx,\\ 
\int V(x)(u_j^{\infty})^2\,dx &\le \liminf_{k\to\infty}\int V(x)(u_j^{n(k)})^2\,dx,
\end{align*}
and 
\begin{align*}
\int f(u_1^{\infty},u_0^{\infty},u_{-1}^{\infty})\,dx
\le \liminf_{k\to\infty}\int f(u_1^{n(k)},u_0^{n(k)},u_{-1}^{n(k)})\,dx
\end{align*}
for every continuous function $f:\mb{R}^3\to[0,\infty)$. 
Therefore, every parts of $\ml{E}$ satisfies such a lower-semicontinuity inequality. 
We claim that these $\liminf$'s are all limits and the inequalities are all equalities provided $\ml{E}[\mf{u}^{n(k)}]\to E_g$. This is obvious, since otherwise we have
\begin{align*}
E_g = \lim_{k\to\infty}\ml{E}[\mf{u}^{n(k)}] 
    > \ml{E}[\mf{u}^{\infty}],
\end{align*}
contradicting to the fact that $\mf{u}^{\infty}\in\mb{A}_M$. 
Now from $\int|\na u_j^{n(k)}|^2\,dx\to\int|\na u_j^{\infty}|^2\,dx$
and (\ref{cofnorm}), 
\begin{align}\label{convofnorm}
\|u_j^{n(k)}\|_{H^1}\to\|u_j^{\infty}\|_{H^1}\quad (j=1,0,-1).
\end{align}
Since $H^1(\mb R^3)$ is reflexive, (\ref{convofnorm}) together with the fact 
$\mf{u}^{n(k)}\wto\mf{u}^{\infty}$ weakly in $H^1(\mb R^3)$ implies 
$\mf{u}^{n(k)}\to\mf{u}^{\infty}$ strongly in $H^1(\mb R^3)$. 
Similarly we can prove $\mf{u}^{n(k)}\to\mf{u}^{\infty}$ in 
$L_V^2(\mb R^3)$ and in $L^4(\mb R^3)$, and hence
$\mf{u}^{n(k)}\to\mf{u}^{\infty}$ in the norm of $\mb{B}$. 
This completes the proof.

\bibliographystyle{plain}
\bibliography{bec}
\end{document}